\documentclass[a4paper]{jpconf}
\usepackage{graphicx}

\newcommand\ro{{\hat{\rho}}}
\newcommand\Ho{\hat H}
\newcommand\fo{\hat f}
\newcommand\Ao{\hat A}
\newcommand\Herm{\mathrm{Herm}}
\begin{document}
\title{The gravity-related decoherence master equation from hybrid dynamics}

\author{Lajos Di\'osi}

\address{Research Institute for Particle and Nuclear Physics, H-1525 Budapest 114, P.O.Box 49, Hungary}

\ead{diosi@rmki.kfki.hu}

\begin{abstract}
Canonical coupling between classical and quantum systems cannot result 
in reversible equations, rather it leads to irreversible master
equations. Coupling of quantized non-relativistic matter to gravity is 
illustrated by a simplistic example. The heuristic derivation yields
the theory of gravity-related decoherence proposed longtime ago by 
Penrose and the author. 
\end{abstract}

\section{Introduction}
There is at least one classical dynamic system whose quantization is still 
problematic, that is gravity. Experimental evidences up to now 
have not indicated that gravitation should be quantized. Methods 
of quantizing the classical equations of gravity have failed to become
generally accepted. So, gravitation could happen to be classical. 
Classical dynamics of gravitation would couple to quantum dynamics of other fields. 
This mathematical problem is not at all trivial. 

Hybrid quantum-classical systems are important not only for gravity but for molecular 
and nuclear systems as well where we might need to go beyond the mean-field approximation.
Hybrid dynamics has obtained certain theoretical importance in foundations, in cosmology,    
in measurement problem. A very incomplete list of related works 
\cite{Ale,BouTra,Dio95,Sal,CarSal,DioGisStr,PreKis,Dia,PerTer,Sah,Kis,RegHal}
shows the diversity of motivations.

The particular goal of the present work
is to re-derive a former heuristic gravity-related decoherence model. According
to it, the Newton gravitational field $\phi$ possesses a universal uncertainty $\delta\phi$
which is modelled by a zero-mean white-noise with spatial 
correlation \cite{DioLuk,Dio87}
\begin{equation}\label{Corr87}
\left\langle
\delta \phi(r,t^\prime)\delta \phi(s,t)\right\rangle_{noise}
        =\frac{G\hbar}{\vert r-s\vert}\delta(t^\prime-t)\;,
\end{equation} 
where $G$ is the Newton constant.
This noise causes a certain universal decoherence between 'macroscopically'
different mass distributions. The following quantum master equation
can be derived \cite{Dio87,Dio89} for the density matrix $\ro_Q$ of the quantized
system:
\begin{equation}\label{mast87}
\frac{d}{dt}{\ro}_Q=-{i\over\hbar}[\Ho_Q+\Ho_G,\ro_Q]
-{1\over2}\int_{r,s}{G/\hbar\over\vert r-s \vert}
        [\fo(r),[\fo(s),\ro_Q]\;,
\end{equation}
where $\Ho_Q$ is the Hamiltonian, $\Ho_G$ is the standard Newton pair-potential,
$\fo(r)$ is the operator of spatial mass density. 
(We have omitted the integral volume elements $dr,ds$ and we keep this
shorthand notation throughout our work.) 
The decoherence mechanism
encoded in Eqs.~(\ref{Corr87},\ref{mast87}) coincides (upto a missing factor $1/2$) with Penrose's proposal \cite{Pen94,Pen96}  
\begin{equation}\label{Pen}
\frac{1}{4}\int_{r,s}{G/\hbar\over\vert r-s \vert}[f(r)-f'(r)][f(s)-f'(s)]
\end{equation}
for the rate of decoherence between two different mass distributions $f$ and $f'$.

Sec.~\ref{Hybrid_dynamics} introduces the mathematical model of hybrid dynamics
which will be applied to the matter-gravity quantum-classical hybrid system in 
Sec.~\ref{Gravity_related_decoherence} including a re-derivation of Eqs.~(\ref{Corr87},\ref{mast87}).

\section{Hybrid dynamics}\label{Hybrid_dynamics}
We are going to discuss a possible mathematical model for hybrid composite systems
consisting of one quantum and one classical subsystem. 

\subsection{Hybrid state, expectation values}\label{Hybrid_state_expectation_values}
The generic state of a hybrid system is the hybrid density 
\begin{equation}\label{ro}
\ro\equiv\ro(q,p)
\end{equation}
which is a non-negative matrix in the Hilbert space of the quantum subsystem, depending
on the phase space point $(q,p)$ of the classical subsystem. Its quantum marginal
\begin{equation}\label{ro_Q}
\ro_Q\equiv\int\ro(q,p)dqdp
\end{equation}
is the usual normalized density matrix of the quantum subsystem, while its classical marginal
\begin{equation}\label{rho_C}
\rho_C(q,p)\equiv\mathrm{tr}\ro(q,p)
\end{equation}
is the usual normalized Liouville density of the classical subsystem. (We mention
the notion of conditional state $\ro_Q\vert_{qp}\equiv\ro(q,p)/\rho_C(q,p)$ of the quantum subsystem;
conditional state of the classical subsystem wouldn't make sense in general.) 

The interpretation of the
hybrid state follows from the interpretation of density matrices and Liouville densities,
and goes like this. Consider an arbitrary phase-point-dependent Hermitian matrix $\Ao(q,p)$,
i.e.: a hybrid observable. Its expectation value in the hybrid state $\ro$ is defined by
\begin{equation}\label{expval} 
\langle\Ao(q,p)\rangle_\ro\equiv\mathrm{tr}\int\Ao(q,p)\ro(q,p)dqdp\;.
\end{equation}
The expectation value has the usual statistical interpretation: if we measure the value
of the hybrid observable $\Ao(q,p)$ repeatedly on a large ensemble of identically
prepared hybrid systems characterized by the hybrid state $\ro$, the average value of
the measured outcomes will tend to the above calculated expectation value.

Once we have specified the general features of hybrid states, we can go on and
learn about their dynamics. 
   
\subsection{Dirac, Poisson, Aleksandrov brackets}
Let us begin with two independent systems, one is quantum and the other is classical. 
Our quantum system's state is described by 
the density matrix $\ro_Q$. The evolution is governed by the  
von Neumann equation with Hamiltonian $\Ho_Q$:
\begin{equation}\label{vNeu}
\frac{d}{dt}{\ro}_Q=-\frac{i}{\hbar}[\Ho_Q,\ro_Q]\equiv -{i\over\hbar}\left(\Ho_Q\ro_Q-\ro_Q\Ho_Q\right)\;.
\end{equation}
For our classical system, we have chosen the classical canonical Liouville theory
because of its best formal match with the quantum theory. The state of the classical system
is described by the normalized phase space density $\rho_C=\rho_C(q,p)$. The evolution is governed
by the Liouville equation with Hamilton function $H_C=H_C(q,p)$: 
\begin{equation}\label{Liou}
\frac{d}{dt}\rho_C=\{H_C,\rho_C\}_P\equiv \sum_n 
\left({\partial    H_C\over\partial q_n}{\partial \rho_C\over\partial p_n}
     -{\partial \rho_C\over\partial q_n}{\partial    H_C\over\partial p_n}\right)\;,
\end{equation}
where $q=(q_1,q_2,\dots,q_n)$ and $p=(p_1,p_2,\dots,p_n)$ stand for the
canonical pairs of coordinates and momenta.
The classical structure $\{\;,\;\}_P$ is called Poisson bracket, 
the quantum structure $-(i/\hbar)[\;,\;]$ is called Dirac bracket.

Now we can take the above quantum and classical systems to construct a 
single composite hybrid system. We can easily form the hybrid composite state: 
\begin{equation}\label{rohyb0}
\ro=\ro(q,p)\equiv\ro_Q\rho_C(q,p)\;,
\end{equation}
as well as the hybrid Hamiltonian: 
\begin{equation}\label{Hhyb0}
\Ho=\Ho(q,p)\equiv\Ho_Q+H_C(q,p)\;.
\end{equation}
We write the hybrid evolution equation in the form
\begin{equation}\label{Ale0}
\frac{d}{dt}{\ro}=-\frac{i}{\hbar}[\Ho,\ro]+\{\Ho,\ro\}_P\;,
\end{equation}
which is equivalent with the previous evolution equations (\ref{vNeu}) and (\ref{Liou}) for
$\ro_Q$ and $\rho_C$ separately. 

Of course, we would like to introduce interaction between the quantum and classical
subsystems, i.e., we introduce the hybrid interaction Hamiltonian $\Ho_{QC}=\Ho_{QC}(q,p)$
which is a certain Hermitian matrix that also depends on the canonical coordinates
of the classical subsystem. Now our total hybrid Hamiltonian reads 
\begin{equation}\label{Hhyb}
\Ho=\Ho(q,p)\equiv\Ho_Q+H_C(q,p)+\Ho_{QC}(q,p)\;,
\end{equation}
and Aleksandrov \cite{Ale} proposed the following evolution equation:
\begin{equation}\label{Ale}
\frac{d}{dt}{\ro}=-\frac{i}{\hbar}[\Ho,\ro]+\Herm\{\Ho,\ro\}_P\;.
\end{equation}
The structure on the r.h.s. is called the Aleksandrov bracket, being the
simplest generalization of the naive interaction-free structure (\ref{Ale0}).

The Aleksandrov equation (\ref{Ale}) does not preserve the factorized form (\ref{rohyb0}) of the
hybrid state $\ro(q,p)$, which is the legitimate consequence of the interaction. But a serious drawback arises. 
As pointed out by Boucher and Traschen \cite{BouTra}, the positivity of $\ro(q,p)$ 
may not be preserved, the Aleksandrov equation is fundamentally defective whereas it
can be a useful approximation beyond the mean-field model.

\subsection{Blurring Dirac+Poisson}
However disappointing it is, the algebraic unification of classical and quantum 
dynamics runs into difficulties \cite{BouTra,Dio95,Sal,CarSal}. Perhaps the deepest reason is that the notion of 
deterministic classical trajectories of one subsystem 
becomes lost under the influence of the other subsystem which is subject to quantum
uncertainties. I proposed a possible remedy long time ago \cite{Dio95}, another 
approach was shown together with Gisin and Strunz \cite{DioGisStr}; 
further mathematical structures of hybrid dynamics appear from time to time 
\cite{PreKis,Dia,PerTer,Sah,Kis,RegHal}.  
A comparative analysis is missing. The present work follows the method
of Ref.~\cite{Dio95}. 

In the hybrid Hamiltonian (\ref{Hhyb}), the interaction can always be
written in the form 
\begin{equation}\label{HQCJJ}
\Ho_{QC}=\sum_r\hat J_Q^r J_C^r\;,
\end{equation}
where Ref.~\cite{Dio95} called $\hat J_Q^r$ and $J_C^r$ as `currents' that build up
the hybrid coupling. This time, with one eye 
on the forthcoming application in Sec.~\ref{Gravity_related_decoherence}, 
we write the hybrid interaction this way:     
\begin{equation}\label{HQCfphi}
\Ho_{QC}=\sum_r \fo^r \phi^r\;.
\end{equation}
Let us blur the interacting 'currents' by auxiliary
classical noises $\delta f^r$ and $\delta \phi^r$: 
\begin{equation}\label{HQCnoise}
\Ho_{QC}^{noise}=\sum_r 
\left(\fo^r+\delta f^r(t)\right)\left(\phi^r+\delta \phi^r(t)\right)\;.
\end{equation}
Accordingly, we replace the hybrid Hamiltonian (\ref{Hhyb}) by the noisy one:
\begin{equation}\label{Hhybnoise}
\Ho^{noise}=\Ho_Q+H_C+\Ho_{QC}^{noise}\;,
\end{equation}
and we replace the Aleksandrov equation (\ref{Ale}) by its noise-averaged (blurred, coarse-grained) version:
\begin{equation}\label{Ale_ave}
\frac{d}{dt}{\ro}=\left\langle-\frac{i}{\hbar}[\Ho^{noise},\ro]+\Herm\{\Ho^{noise},\ro\}_P\right\rangle_{noise}\;.
\end{equation}
Suppose the noises $\delta f$ and $\delta\phi$ are independent white-noises of zero mean, 
with the following auto-correlations:
\begin{eqnarray}\label{Corrff}
\left\langle
\delta f^r(t^\prime)\delta f^s(t)\right\rangle_{noise}
        \!\!\!&=&\!\!\!D_Q^{rs}\delta(t^\prime-t)\;,
\\
\label{Corrphiphi}
\left\langle
\delta \phi^r(t^\prime)\delta \phi^s(t)\right\rangle_{noise}
        \!\!\!&=&\!\!\!D_C^{rs}\delta(t^\prime-t)\;.
\end{eqnarray}
The two correlation matrices $D_Q^{rs}$ and $D_C^{rs}$ can be chosen in such
a way that the noise-averaged Aleksandrov equation (\ref{Ale_ave}) guarantee the
positivity of the coarse-grained hybrid density $\ro$. The price to pay is the loss
of reversibility: the mathematically correct hybrid equation will be a certain
irreversible master (kinetic) equation. 

\subsection{Hybrid master equation}
By hand, we added a noisy part (\ref{HQCnoise}) to the total hybrid Hamiltonian (\ref{Hhyb}):
\begin{equation}\label{Hnoise}
\Ho^{noise}=\Ho+\sum_r (\fo^r\delta \phi^r+\phi^r\delta f^r)\;.
\end{equation}
Compared to Eqs.~(\ref{HQCnoise},\ref{Hhybnoise}), the term $\delta f\delta \phi$ has been omitted because it cancels from both the
Dirac and Poisson brackets. In the evolution equation (\ref{Ale_ave}),
the $1^{st}$ order contribution of the noisy part is vanishing on average. In $2^{nd}$ order,
the term $\fo\delta\phi$ adds the structure $-[\fo,[\fo,\ro]]$ and 
the term $\phi\delta f$ adds the structure $\{\phi,\{\phi,\ro\}_P\}_P$ to the r.h.s. of the naive
Aleksandrov equation (\ref{Ale}): 
\begin{equation}\label{hybmast}
\frac{d}{dt}{\ro}=-{i\over\hbar}[\Ho,\ro]+\Herm\{\Ho,\ro\}_P
-\frac{1}{2\hbar^2}\sum_{r,s}
       D_C^{rs}[\fo^r,[\fo^s,\ro]]
+\frac{1}{2}\sum_{r,s}
       D_Q^{rs}\{\phi^r,\{\phi^s,\ro\}_P\}_P\;.
\end{equation}
This hybrid master equation, i.e.: the noise-averaged Aleksandrov equation (\ref{Ale_ave}), 
preserves the positivity of $\ro$ if the correlation matrices
satisfy
\begin{equation}\label{DQDC}
D_QD_C\geq\frac{\hbar^2}{4}\;. 
\end{equation}
The proof was given in Ref.~\cite{Dio95} (it has remained largely unnoticed).

The first irreversible term on the r.h.s. of the hybrid master equation (\ref{hybmast}) 
is a typical decoherence term dephasing the
superpositions between different configurations of the 'currents' $f$. The kernel $D_C$ plays the
role of decoherence matrix. The second irreversible term concerns the classical subsystem,
yielding diffusion in its phase space. 

\section{Gravity-related decoherence}\label{Gravity_related_decoherence}
Hybrid dynamics will be applied to the interaction of quantized massive particles with
classical gravitational field.  

\subsection{Hybrid master equation for matter plus gravity}
The Hamiltonian $\Ho_Q$ will stand for the quantized non-relativistic matter. 
We'll need its mass density field operator $\fo$, which should be the non-relativistic 
limit of ${\hat T}_{00}/c^2$, the $00$ component of the Einstein energy-momentum
tensor divided by $c^2$ ($c$ is the speed of light). For point-like particles 
of mass $m$ and position operator $\hat x$, the mass density would be 
\begin{equation}\label{fr}
\fo(r)=\sum m\delta(r-\hat x)\;, 
\end{equation}
where summation extends for all massive particles. In fact, 
a certain short-length cutoff is always understood, for details 
of this issue see Ref.~\cite{Dio05,Dio07a,Dio07b} and references therein.  

As to the classical gravitational field, we start with a scalar relativistic model 
and we take the non-relativistic limit $c\rightarrow\infty$ afterwards. 
We define the Newton potential $\phi\equiv{1\over2}c^2(g_{00}-1)$ 
where $g_{00}$ is the $00$ component of the Einstein metric tensor. 
The canonical coordinates $q$ of Sec.~\ref{Hybrid_dynamics} will be played by
the Newton field $\phi(r)$, the conjugate canonical momenta $p$ will be denoted 
by the field $\xi(r)$. The Hamilton function(al) can be chosen as 
\begin{equation}\label{HCphixi}
H_C(\phi,\xi)=
\int_r\left(2\pi Gc^2~\xi^2+\frac{\vert\nabla\phi\vert^2}{8\pi G}\right)\;.
\end{equation}
To construct the hybrid interaction, we couple the 
quantized mass density $\fo$ to the Newton field: 
\begin{equation}\label{HQCgrav}
\Ho_{QC}(\phi)=\int_r \fo(r)\phi(r)\;.
\end{equation}
Note, in parentheses, that $\Ho_{QC}$ does not depend on the canonical momenta $\xi$.
Looking back to notations in Sec.~\ref{Hybrid_dynamics}, we recognize the identity 
of this interaction with (\ref{HQCfphi}) provided we replace $\sum_r,\fo^r,\phi^r$ by
$\int_r,\fo(r),\phi(r)$, respectively. 
Therefore the equations of Sec.~\ref{Hybrid_dynamics} apply here conveniently. 
Let us form the total hybrid Hamiltonian (\ref{Hhyb}):
\begin{equation}\label{Hgrav}
\Ho(\phi,\xi)=\Ho_Q+H_C(\phi,\xi)+\Ho_{QC}(\phi)\;.   
\end{equation}
Write down the hybrid master equation (\ref{hybmast}) for the state $\ro(\phi,\xi)$ of our hybrid system: 
\begin{eqnarray}\label{hybmastgrav}
\frac{d}{dt}{\ro}=\!\!\!\!\!\!\!\!&&-{i\over\hbar}[\Ho,\ro]+\Herm\{\Ho,\ro\}_P\nonumber\\
\!\!\!\!\!\!\!\!&&-\frac{1}{2\hbar^2}\int_{r,s}\!\!\!
       D_C(r,s)[\fo(r),[\fo(s),\ro]]
+\frac{1}{2}\int_{r,s}\!\!\!
       D_Q(r,s)\{\phi(r),\{\phi(s),\ro\}_P\}_P\;.
\end{eqnarray}
For preserving the positivity of the state $\ro$, the decoherence and diffusion 
kernels should satisfy the condition (\ref{DQDC}). Below, we propose a heuristic 
resolution of the remaining ambiguity as to the concrete form of $D_Q$ and $D_C$.

\subsection{The gravity-related decoherence matrix}
In the Newtonian non-relativistic limit $c\rightarrow\infty$, the mean-field Poisson equation is valid.
In our notations (\ref{expval}):     
\begin{equation}\label{MF}
\Delta\left<\phi(r)\right>_\ro=4\pi G\left<\fo(r)\right>_\ro\;.
\end{equation}
Let us recall the definitions (\ref{Corrff}) and (\ref{Corrphiphi}) of the correlation 
matrices (kernels) $D_Q$ and $D_C$, respectively. 
If we imposed the Poisson equation $\Delta\delta\phi=4\pi G\delta f$ 
on the fluctuations $\delta f,\delta\phi$ 
in Eqs.~(\ref{Corrff},\ref{Corrphiphi}), respectively, then 
the above two correlation kernels would become related by
\begin{equation}
\Delta\Delta^\prime D_C(r,r^\prime)=(4\pi G)^2D_Q(r,r^\prime)\;.
\end{equation}
(We are aware of a conflict with the derivation of the hybrid master equation where we
considered $\delta\phi$ and $\delta f$ uncorrelated. This is conceptual problem,
it remains harmless for our resulting hybrid master equation but it may become
relevant for later developments.) 
From the inequality (\ref{DQDC}), the minimum blurring corresponds to $D_CD_Q=\hbar^2/4$.
Hence the unique translation invariant solution reads 
\begin{equation}\label{DC}
D_C(r,s)=\frac{G\hbar}{2}\frac{1}{\vert r-s \vert}\;.
\end{equation} 
This is the correlation (\ref{Corr87}) of universal gravitational fluctuations
\cite{DioLuk,Dio87} apart from the numeric factor $1/2$.   

\subsection{Reduced quantum master equation}
Let us invoke the expansion (\ref{Liou}) of the Poisson bracket and 
calculate the full expansion of the r.h.s. of the hybrid master equation
(\ref{hybmastgrav}):
\begin{eqnarray}\label{hybmastgravfull}
\frac{d}{dt}{\ro}=\!\!\!\!\!\!\!\!&&-{i\over\hbar}[\Ho_Q,\ro]
                    -{i\over\hbar}\int_r \phi(r)[\fo(r),\ro]
\nonumber\\
\!\!\!\!\!\!\!\!&&-\int_r\Bigl( 
        4\pi Gc^2\xi(r){\delta\ro\over\delta\phi(r)}
        +{1\over4\pi G}\Delta\phi(r){\delta\ro\over\delta\xi(r)}\Bigr)
+\Herm\int_r \fo(r){\delta\ro\over\delta\xi(r)}
\nonumber\\
\!\!\!\!\!\!\!\!&&-\frac{1}{2\hbar^2}\int_{r,s}\!\!D_C(r,s)
        [\fo(r),[\fo(s),\ro]~]
+\frac{1}{2}\int_{r,s}\!\!D_Q(r,s)
        {\delta^2\ro\over\delta\xi(r)\delta\xi(s)}\;.
\end{eqnarray}
We are going to derive the reduced dynamics of the quantized matter. 
Its density matrix is the marginal (\ref{ro_Q}) of the hybrid state:
\begin{equation}\label{roQ}
\ro_Q=\int\ro(\phi,\xi){\cal D}\phi{\cal D}\xi\;.
\end{equation}
Let us try to obtain a closed master equation for $\ro_Q$.
Integrate both sides of (\ref{hybmastgravfull}) by ${\cal D}\phi$ and ${\cal D}\xi$.
Most terms on the r.h.s. cancel and we are left with
\begin{equation}\label{mast0}
\frac{d}{dt}{\ro}_Q=-{i\over\hbar}[\Ho_Q,\ro_Q]-{i\over\hbar}\int_r\!\!\!\phi(r)[\fo(r),\ro(\phi)]{\cal D}\phi
-\frac{1}{2\hbar^2}\!\!\int_{r,s}\!\!\!\!\!\!
       D_C(r,s)[\fo(r),[\fo(s),\ro]]\;,
\end{equation}
where $\ro(\phi)=\int\ro(\phi,\xi){\cal D}\xi$. Suppose the 
post-mean-field Ansatz:
\begin{equation}\label{PMF}
\int\Delta\phi(r)\ro(\phi){\cal D}\phi=4\pi G~\Herm\fo(r)\ro_Q\;,
\end{equation}
which is stronger than the standard mean-field equation (\ref{MF}).
From this Ansatz, we express the functional integral on the r.h.s. of (\ref{mast0}):
\begin{equation}
\int\phi(r)\ro(\phi){\cal D}\phi=-G\Herm\int_s\frac{\fo(s)}{\vert r-s\vert}\ro_Q\;.
\end{equation}
Subsequently, the complete term in question becomes Hamiltonian:
\begin{equation}
-{i\over\hbar}\int_r\!\!\phi(r)[\fo(r),\ro(\phi)]{\cal D}\phi=-{i\over\hbar}[\Ho_G,\ro_Q]\;,
\end{equation}
where $\Ho_G$ is the well-known Newtonian pair-potential:
\begin{equation}
\Ho_G=-{G\over2}\int_{r,s}{\fo(r)\fo(s)\over\vert r-s\vert}\;.
\end{equation}

The Eq.~(\ref{mast0})
has thus led to the following closed master equation for the evolution
of the quantized matter:
\begin{equation}\label{mast}
\frac{d}{dt}{\ro}_Q=-{i\over\hbar}[\Ho_Q+\Ho_G,\ro_Q]
-\frac{1}{2\hbar^2}\int_{r,s}\!\!\!D_C(r,s)
        [\fo(r),[\fo(s),\ro_Q]\;,
\end{equation}
which, invoking $D_C$ from (\ref{DC}), is the master equation (\ref{mast87}) \cite{Dio87,Dio89}
apart from an additional numeric factor $1/2$ (cf. also Ref.~\cite{Dio09}) in front of the 
decoherence term .

\section{Concluding remarks}\label{Concluding_remarks}
We have constructed a possible consistent hybrid dynamics coupling 
classical gravity to quantized non-relativistic matter. We were able
to re-derive the former heuristic equations of gravity-related universal decoherence.

We admit a certain elusiveness of our procedure. In order to borrow
self-dynamics to the Newtonian field $\phi$, we attributed the status of relativistic
scalar field to $\phi$, with the non-relativistic limit to be taken on the hybrid
master equation. On the other hand, the proposed hybrid master equation assumes the 
Markovian approximation since it contains classical white-noise fields. Coexistence
of relativistic and white-noise fields may be an issue in itself. Therefore our
hybrid master equation is heuristic rather than exact, and it remains so until, e.g., we
construct some explicit solutions at least. The non-relativistic limit $c\rightarrow\infty$
was treated formally, without check for its existence. This is why we keep talking about 
the post-mean-field Ansatz rather than the post-mean-field equation despite the heuristic 
proof presented in the Appendix. 

In any case, the formal re-derivation of the gravity-related decoherence master equation 
from a certain closed set of equations is reassuring and indicates the possible relevance
of hybrid dynamics in foundations \cite{Dio00a,Dio00b}.

\ack
This work was supported by the Hungarian Scientific Research Fund under Grant No. 75129.

\appendix
\section*{Appendix}
We ought to show that the post-mean-field Ansatz (\ref{PMF}) is
consistent with the non-relativistic limit $c\rightarrow\infty$ 
of the hybrid master equation (\ref{hybmastgrav}).
We begin with the heuristic derivation of the standard mean-field equation (\ref{MF}).

From the hybrid master equation (\ref{hybmastgravfull}) and the definition of expectation
values (\ref{expval}),
calculate the time-derivatives of the classical mean fields $\langle\phi\rangle_\ro$ and  
$\langle\xi\rangle_\ro$:
\begin{eqnarray}
\frac{d}{dt}\langle\phi(r)\rangle_\ro\!\!\!&=&\!\!\!      4\pi Gc^2\langle\xi(r)\rangle_\ro\;,
\\
\frac{d}{dt}\langle \xi(r)\rangle_\ro\!\!\!&=&\!\!\!-{1\over4\pi G}\langle\Delta\phi(r)-4\pi G\fo(r)\rangle_\ro\;.
\end{eqnarray}
From the first equation wee see that $\langle\xi\rangle_\ro\rightarrow0$ in the non-relativistic limit.
It also means that the l.h.s. of the second equation vanishes, hence we have:
\begin{equation}
\langle\Delta\phi(r)-4\pi G\fo(r)\rangle_\ro=0\;.
\end{equation}
This is just the standard mean-field equation (\ref{MF}). 

Let us upgrade it into the post-mean-field Ansatz (\ref{PMF}). 
To this end, we calculate the time-derivative of the expectation value of
the simple hybrid quantity $\Ao\phi$ where $\Ao$ itself is an arbitrary constant Hermitian matrix:
\begin{eqnarray}
\!\!\!\frac{d}{dt}\langle\Ao\phi(r)\rangle_\ro\!\!\!\!\!&=&\!\!\!\!4\pi Gc^2\!\langle\Ao\xi(r)\rangle_\ro
+\!{i\over\hbar}\Bigl\langle[
\Ho_Q+\!\!\!\int_s\!\!\!\phi(s)\fo(s),\Ao\phi(r)]-\frac{1}{2\hbar^2}\!\!\!\int_{r,s}\!\!\!\!\!\!D_C(u,s)[\fo(u),[\fo(s),\Ao\phi(r)
                            ]\Bigr\rangle_\ro
\nonumber\\
                                                        &=&\!\!\!\!4\pi Gc^2\langle\Ao\xi(r)\rangle_\ro+{\cal O}(1)\;.
\end{eqnarray}
Here ${\cal O}(1)$ stands for terms that does not contain the diverging factor $c^2$.
In the non-relativistic limit, the quantity $\langle\Ao\xi(r)\rangle_\ro$ must vanish. 
Since $\Ao$ can be any Hermitian operator, the 'partial' expectation value,
i.e.: the expression (\ref{expval}) without the trace and $\Ao$, does already vanish:
\begin{equation}
\int\xi(r)\ro(\phi,\xi){\cal D}\phi{\cal D}\xi=0\;.
\end{equation}
In general, the hybrid dynamics (\ref{hybmastgravfull}) yields
\begin{equation}
\frac{d}{dt}\langle\Ao\phi(s)F(\xi)\rangle_\ro=4\pi Gc^2\langle\Ao\xi(r)F(\xi)\rangle_\ro+{\cal O}(1)\;,
\end{equation}
where $F(\xi)$ can be a generic functional of $\xi$.
Repeating the previous arguments, we conclude that the 'partial' expectation value of 
$\xi(r)F(\xi)$ must vanish for $c\rightarrow\infty$. Since $F[\xi]$ is an arbitrary functional, 
we come to the conclusion that, in the non-relativistic limit, $\xi(r)$ is identically zero. 
Now we derive the time-derivative of $\langle\Ao\xi(r)\rangle_\ro$ from (\ref{hybmastgravfull}):
\begin{eqnarray}
\frac{d}{dt}\langle\Ao\xi(r)\rangle_\ro=
\!\!\!&-&\!\!\!\!\!{1\over4\pi G}\langle\Ao\Delta\phi(r)-4\pi G\Herm\Ao\fo(r)\rangle_\ro
\\
      &+&\!\!\!\!\!{i\over\hbar}
\Bigl\langle[
\Ho_Q+\int_s\!\!\phi(s)\fo(s),\Ao\xi(r)]-\frac{1}{2}\int_{r,s}\!\!\!\!\!D_C(u,s)[\fo(u),[\fo(s),\Ao\xi(r)
]\Bigr\rangle_\ro\;.
\nonumber
\end{eqnarray}
As we said, in the limit $c\rightarrow\infty$ we can set $\xi$ to zero identically for all time and position 
therefore the above equation reduces to:
\begin{equation}
\langle\Ao\Delta\phi(r)-4\pi G\Herm\Ao\fo(r)\rangle_\ro=0\;.
\end{equation}
The l.h.s. vanishes for all $\Ao$, hence we can remove the trace and $\Ao$ from the hybrid expectation value, yielding:
\begin{equation}
\int\left(\Delta\phi(r)-4\pi G\Herm\fo(r)\right)\ro(\phi,\xi){\cal D}\phi{\cal D}\xi=0\;.
\end{equation}
This is exactly the post-mean-field Ansatz (\ref{PMF}) if we invoke the definition (\ref{ro_Q}) of $\ro_Q$.

\end{document}